\documentclass[fleqn%
]
{article}
\usepackage{espcrc2}
\usepackage{epsfig}
\newcommand{\mysection}{\section}

\begin{document} 
 \title{Fullerene based devices for molecular electronics}
 \author{
 {G. Cuniberti}$^a$\thanks{e--mail: \texttt{cunibert@mpipks-dresden.mpg.de}},
 R. Gutierrez$^b$, 
 G. Fagas$^a$, 
 F. Grossmann$^b$, 
 K. Richter$^c$, 
 and R. Schmidt$^b$
\\[.3cm] 
{\small \it
$^a$ Max Planck Institute for the Physics of Complex Systems, 
	 N{\"o}thnitzer Str. 38, D-01187 Dresden
\\ $^b$ Institute for Theoretical Physics, Technical University of Dresden,
D-01062 Dresden
\\ $^c$ Institute for Theoretical Physics, University of Regensburg,
	 D-93053 Regensburg
}
}
\maketitle

\def\c60{C$_{60}$}
\def\rmw{\rm mol}
\def\ie{i.e.$\!$}
\def\eg{e.g.$\!$}
\def\ee{{\rm e}}
\def\ii{{\rm i}}
\setlength{\parindent}{0.pt}

 \newcommand{\vettore}{\vec}
 \newcommand{\matrice}[1] {{\mathbf{#1}}}
 \newcommand{\overlap} {{\cal O}}
 \newcommand{\calg} {{\cal G}}
 \newcommand{\vt} {\vartheta}
 \newcommand{\bea} {\begin{eqnarray}}
 \newcommand{\eea} {\end{eqnarray}}
 \newcommand{\beann} {\begin{eqnarray*}}
 \newcommand{\eeann} {\end{eqnarray*}}
 \newcommand{\labs} {\left\vert}
 \newcommand{\rabs} {\right\vert}
 \newcommand{\lsb} {\left[}
 \newcommand{\rsb} {\right]}
 \newcommand{\lrb} {\left(}
 \newcommand{\rrb} {\right)}
 \newcommand{\lcb} {\left\{}
 \newcommand{\rcb} {\right\}}
 \newcommand{\lab} {\left\langle}
 \newcommand{\rab} {\right\rangle}
 \newcommand{\ve} {\varepsilon}
 \newcommand{\ds} {\displaystyle}
\section*{Abstract}\label{sect:abs}
 We have investigated the electronic properties of a C$_{60}$ molecule in between carbon nanotube leads.  This problem has been tackled within a quantum chemical treatment  utilizing a density functional theory--based LCAO approach combined with the Landauer formalism.  Owing to low--dimensionality, electron transport is very sensitive to the strength and geometry of interfacial bonds.  Molecular contact between interfacial atoms and electrodes gives rise to a complex conductance dependence on the electron energy exhibiting spectral features of both the molecule and electrodes. These are attributed to the electronic structure of the C$_{60}$ molecule and to the local density of states of the leads, respectively. 
\\ {July 23, 2001}

\mysection{Introduction}\label{sect:intro}

The accelerated down-scaling of electronic devices has reached the 
single molecule domain.
As a consequence the investigation of the mechanism with which a single molecule carries
an electric current becomes crucial in view of the possible exploitation
of molecular electronic circuits.
Indeed scanning tunneling microscope (STM) setups and molecular break junctions have already provided
new experimental data concerning transport through individual molecules. 
The selection of the bridge--molecule and the accurate controls for checking that a very single molecule is finally trapped between two electrodes are basic prerequisites for the construction of single molecule electronic devices. A benzene ring was among the first bridge--molecules~\cite{RZMBT97}, and recently also heavier molecules as \c60 have been studied in a break junction configuration~\cite{PPLAAMcE00} and by means of STM~\cite{PGHRBSPAOS00,HJHQCHWCQ99,PLTM97,JSGC95}. 

\begin{figure}[t]
 \centerline{\epsfig{file=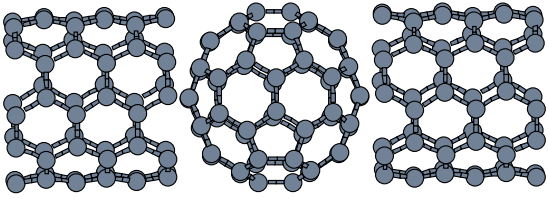, width=.99\linewidth}}
 \caption{\label{fig:sketchc60} The (5,5)--\c60--(5,5) carbon hybrid.}
 \end{figure}

A great concern was also directed to the characterization of the nature of the electrodes and the quality of the contacts with the molecule. In recent experiments, the resolution of STM tips have been enhanced by attaching to them carbon nanotubes (CNTs) segments~\cite{WMSS01,NKANHYT00,HCL99,WJWCL98}.
This gives support to the idea that CNTs can indeed be employed as wiring elements in molecular circuits~\cite{RKJTCL00,HOYL99,BHTSSBF98}. 
In this paper, we will show the results obtained for the conductance through a structure consisting of a single \c60 molecule grasped between two armchair (5,5) CNTs.
(a sketch of the device is illustrated in Figure~\ref{fig:sketchc60}.). 
This design is the natural evolution of a CNT hybrid structure that has been introduced in previous works where linear molecules (molecular wires) have been considered at the tight--binding level~\cite{FCR01a,CFR01a,CFR01b}. 
Here, the description of the hybrid is obtained at a density functional
theory (DFT) level which has been successfully applied to the study of the
conductance through small sodium clusters~\cite{GGKS01}.
%
\mysection{System and Method}\label{sect:sys_meth}
In order to derive transport properties, we make use of the Landauer theory~\cite{IL99} which relates the conductance of the system to an independent--electron scattering problem~\cite{FG99}.
The electron wavefunction is assumed to extend coherently across the device and the two--terminal, linear--response conductance at zero temperature, $g$, is simply proportional to the total transmittance for injected electrons $T(E_{\rm F})$ at the Fermi energy $E_{\rm F}$:
\begin{equation}
\label{eq:linear_zero_temperature_conductance}
g = \frac {2 e^2} h T(E_{\rm F}).
\end{equation}
The factor two accounts for spin degeneracy. 
The transmission function can be calculated from the knowledge of the molecular energy levels, the nature and the geometry of the contacts. 
It is given by 
\bea
T(E)=
\sum_{j_{\rm L},j_{\rm R}} \labs S_{j_{\rm L} j_{\rm R}} \rabs^2=
{\rm Tr} \lcb \matrice{S} \matrice{S}^\dagger \rcb,
\label{eq:transmission_function}
\eea
where $j_{\rm L},j_{\rm R}$ are quantum numbers labelling open channels for transport which belong to mutually exclusive leads, in our case the two semi--infinite perfect nanotubes. The attached molecular system acts as a scatterer, and $\matrice{S}$ is the corresponding quantum--mechanical scattering matrix. The quantity $\labs S_{j_{\rm L} j_{\rm R}}\rabs^2$ is the probability that a carrier coming from, say, left of the scatterer in the transversal mode $j_{\rm L}$ will be transmitted to the right in the transversal mode $j_{\rm R}$. The sum in~(\ref{eq:transmission_function}) is restricted to transversal modes whose energy is smaller than $E_{\rm F}$.


To calculate the transmission, one can 
write down the Green function matrix of the
``extended'' molecule
$\matrice{\calg}^{-1} = \matrice{\calg}_{\rmw}^{-1} + \matrice{\Sigma}_{\rm L} + \matrice{\Sigma}_{\rm R}$
written in terms of the bare molecule Green function and the self--energy
correction due to the presence of the leads. Making use of the Fisher--Lee
relation~\cite{FL81} one can finally write 
\bea
\label{eq:transm_mit_spectral_densities}
T(E)= 4 {\rm Tr } \lcb 
\matrice{\Delta}_{\rm L}
\matrice{\calg}^\dagger 
\matrice{\Delta}_{\rm R}
\matrice{\calg} 
\rcb ,
\eea
where 
\beann
\matrice{\Delta}^{\phantom{\dagger}}_{\alpha} (E) = \frac \ii 2 \left . {\lrb {\matrice{\Sigma}}^{\phantom{\dagger}}_\alpha (z) -{\matrice{\Sigma}}_\alpha^\dagger (z)\rrb }\rabs_{z=E + \ii 0^+},
\eeann
and
the self--energy matrices ${\matrice{\Sigma}}_\alpha$ account for the contact of the molecule to the CNT leads:
\bea
\label{eq:self_en_matr}
{\matrice{\Sigma}}_\alpha = 
\matrice{\Gamma}^\dagger_\alpha \matrice{G}^{\phantom{\dagger}}_{\alpha} \matrice{\Gamma}^{\phantom{\dagger}}_\alpha .
\eea
Here, $\matrice{\Gamma}_{\alpha}$ is the coupling between the molecule and $\alpha$--lead. $\matrice{G}_\alpha$ is the Green function of the semi--infinite $\alpha$--CNT. The coupling matrices are short--range so that they mainly couple
the C$_{60}$ to the first unit cell of the nanotube. Thus, $\matrice{G}_\alpha$
becomes a surface Green's function which has been calculated using the
decimation procedure of L{\'o}pez Sancho {\it et al}~\cite{LSLSR84,LSLSR85}.
The implementation of the introduced transport approach needs as a 
further step the characterization of the hamiltonian, 
and the calculation of the coupling
$\matrice{\Gamma}_{\rm L,R}= \matrice{V}_{\rm L,R} - E \matrice{\overlap}_{\rm L,R}$. Here, in addition to the hamiltonian matrix elements $\matrice{V}_{\rm L,R}$ one has to take into account the non--orthogonal contributions in the orbital basis  that may result from 
the implemented method via the  overlap matrix $\matrice{\overlap}_{\rm L,R}$ between the C$_{60}$ and the left/right lead.
The calculation of $\calg$ and $\Gamma$ has been done by means of an approximate DFT treatment~\cite{SS92,KSS99} based on a linear combination of atomic orbitals (LCAO) ansatz.
There, for the Kohn-Sham electronic single-particle states one gets  
\begin{eqnarray}
\psi_i(\vettore{r}) = \sum_{\mu} c^{(i)}_{\mu} \phi^{\phantom{(i)}}_{\mu}\lrb \vettore{r}-\vettore{R}_{\mu}\rrb ,
\end{eqnarray}
where $\phi_{\mu}\lrb \vettore{r}-\vettore{R}_{\mu}\rrb$'s  are non-orthogonal 
valence atomic orbitals localized at the ionic positions $\vettore{R}_{\mu}$.
With this {\em ansatz}
the Kohn-Sham equations for $\psi_i$ 
are transformed into a set of algebraic equations
\begin{eqnarray}
 \sum_{\nu} \lrb H_{\mu\nu}-\overlap_{\mu\nu} E_i \rrb  c^{(i)}_{\nu} = 0,
\end{eqnarray}
where
$\overlap_{\mu\nu}= \lab \phi_{\mu}|\phi_{\nu} \rab$, and 
$H_{\mu\nu} = \lab \phi_{\mu}| t + V_{\rm eff} | \phi_{\nu} \rab$ 
are the overlap and Hamiltonian matrix elements, respectively; 
$t$ is the one-electron kinetic energy operator.
The effective potential $V_{\rm eff}$ contains contributions from an external 
potential,
the Coulomb potential and the exchange-correlation potential treated in the 
local-density approximation (LDA). 
It is approximated by a sum of atomic 
contributions. 
This formalism can be used to calculate forces when studying  structure
properties or to provide the matrix elements that  serve as input for Green function
based transport calculations. Both procedures have 
recently been carried out for the relaxation and conductance 
calculation of sodium clusters~\cite{GGKS01}.  

\mysection{Results and Conclusions}\label{sect:results}

The exposed method has been applied to solve the transport problem of an unrelaxed pure--carbon two--terminal structure.  
Namely, we have considered  a CNT--C$_{60}$--CNT hybrid, 
with open--end (5,5) single--wall CNTs and the C$_{60}$ rigidly blocked in between at a fixed orientation.
The choice of the particular chirality of the metallic tubes is the one with the best match between tube and C$_{60}$ diameters.
As a free parameter we have chosen the tube--tube distance $s$. The coordinates have been implemented in the DFT algorithm for calculating coupling and Green functions.

\begin{figure}[t]
\centerline{\epsfig{file=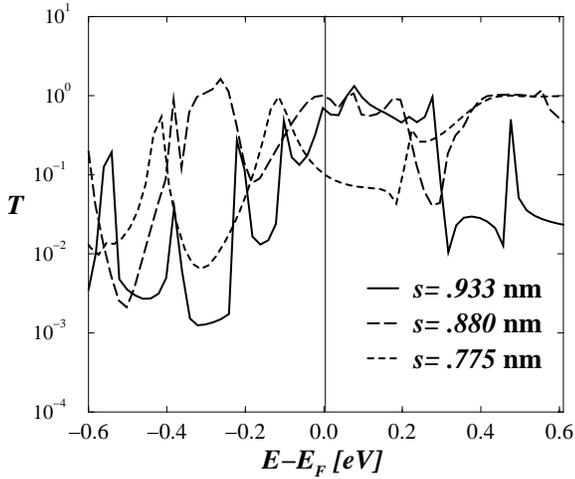, width=\linewidth}}
 \caption{\label{fig:spectrum} Transmission function 
 of the structure for different distances $s$ between the nanotube leads   for
 a fixed orientation of the C$_{60}$.}
 \end{figure}

Typical transmission spectra are plotted in Figure~\ref{fig:spectrum}. Different curves correspond to different distances $s$ between the nanotube leads.

%
%
%
%
%
%

As one can see the conductance shows a great variety of profiles with
differences in magnitude up to three orders. At the Fermi level the
conductance does not seem to follow a monotonic behavior as a function of the
tube--tube distance $s$. The HOMO and LUMO level of the molecule cannot be
easily identified from such conductance profiles. The interaction with the
leads is definitely responsible for their broadening, splitting and shift.

The Fermi level $E_{\rm F}$ has been calculated by considering a
supramolecular structure consisting of the C$_{\rm 60}$ and 6 unit cells in both left and right lead.
Charge transfer is here much less important than in structures with different contacted material -- \eg \ for C$_{60}$ contacted to Al leads~\cite{PPJLV01}.  
We are dealing with an all--carbon structure, this is the reason why the Fermi level lies in the HOMO--LUMO gap of the isolated C$_{\rm 60}$. 


On one hand, these results are extremely comforting when thinking to the
possible effects that the realization of such a device might imply.
In a dual--probe scanning tunneling microscope, similar to the one introduced by Watanabe {\it et al.} in Ref.~\cite{WMSS01}, 
the realization of a CNT--\c60--CNT hybrid 
would be feasible.
But on the other hand this same technique is limited by the fact that the distance between the two tubes could not be rendered smaller than the apex length.

However, a word of caution should be exerted.  
We think that as a next step, the present method should be complemented with a
relaxation procedure in order to verify and control the stability of the
structure~\cite{Gutierrezetal01}.

This research was supported by the ``Deutsche Forschungsgemeinschaft''
through the Forschergruppe ``Nano\-struk\-turier\-te Funk\-tions\-elemente
in makros\-kopi\-schen Sys\-temen''. RG gratefully acknowledges financial
support by the ``S\"achsische Ministerium f\"ur Wissenschaft und Kunst''.
GC research at MPI is sponsored by the Schloe{\ss}mann Foundation.



\end{document}